\documentclass[aps,prd,preprint,nofootinbib]{revtex4}

\usepackage{graphicx}
\usepackage{psfrag}
\usepackage{epsfig}

\begin{document}

\title{A Model for the Twist-3 Wave Function of the Pion and Its
Contribution to the Pion Form Factor}
\author{Tao Huang$^{1,2}$\footnote{email:
huangtao@mail.ihep.ac.cn} and Xing-Gang Wu$^{2}$\footnote{email:
wuxg@mail.ihep.ac.cn}}
\address{$^1$CCAST(World Laboratory), P.O.Box 8730, Beijing 100080,
P.R.China,\\
$^2$Institute of High Energy Physics, Chinese Academy of Sciences,
P.O.Box 918(4), Beijing 100049, China.\footnote{Mailing address}}

\begin{abstract}
A model for the twist-3 wave function $\psi_p(x,\mathbf{k_\perp})$
of the pion has been constructed based on the moment calculation
by applying the QCD sum rules, whose distribution amplitude has a
better end-point behavior than that of the asymptotic one. With
this model wave function, the twist-3 contributions including both
the usual helicity components ($\lambda_1+\lambda_2=0$) and the
higher helicity components ($\lambda_1+\lambda_2=\pm 1$) to the
pion form factor have been studied within the modified pQCD
approach. Our results show that the twist-3 contribution drops
fast and it becomes less than the twist-2 contribution at $Q^2\sim
10GeV^2$. The higher helicity components in the twist-3 wave
function will give an extra suppression to the pion form factor.
The model dependence of the twist-3 contribution to the pion form
factor has been studied by comparing four different models. When
all the power contributions, which include higher order in
$\alpha_s$, higher helicities, higher twists in DA and etc., have
been taken into account, it is expected that the hard
contributions will fit the present
experimental data well at the energy region where pQCD is applicable.\\


\end{abstract}
\maketitle

\section{Introduction}

The most challenging problems for applying the perturbative QCD
(pQCD) to exclusive processes have long been discussed and
analyzed in many papers, such as the pQCD applicability to the
exclusive processes at experimentally accessible energy region due
to the end-point singularity; to estimate the contributions from
power corrections, which includes higher order in $\alpha_s$,
higher helicities, higher twists in distribution amplitude (DA),
higher Fock states and etc.; to estimate the uncertainties from
perturbatively incalculable DAs.

The pion form factor can be obtained through the definition
\begin{equation}
\langle \pi(p^{\prime})|J_{\mu}|\pi(p)\rangle=(p+p^{\prime})_{\mu}
F_{\pi}(Q^2),
\end{equation}
where $J_{\mu}=\sum_i e_i\bar{q}_i\gamma_{\mu}q_{i}$, with the
quark flavor $i$ and the relevant electric charge $e_i$, is the
vector current. The momentum transfer $q^2=-Q^2=(p-p^{\prime})^2$
is restricted in the space-like region. The pQCD applicability to
the pion form factor at the experimentally accessible energy
region has been raised by Ref.\cite{isgur} and attracted much
attention for many years. In the modified pQCD approach that is
proposed in Ref.\cite{lis}, i.e. the transverse momentum
dependence ($k_T$ dependence) as well as the Sudakov corrections
are taken into account in the calculations, we have the following
factorization formula\cite{lis,pi2pi,jps,cchm},
\begin{equation}\label{pikt}
F_{\pi}(Q^2)=\sum_{n,m}\int[dx_id{\bf k}_{\perp
i}]_{n}[dy_jd{\bf l}_{\perp j}]_{m} \psi_{n}(x_i,{\bf k}_{\perp
i};\mu_f)T_{n m}(x_i,{\bf k}_{\perp i};y_j,{\bf l}_{\perp
j};\mu_f)\psi_{m}(y_j,{\bf l}_{\perp j};\mu_f),
\end{equation}
where $[dx_i d\mathbf{k}_{i}]_n$ is the relativistic measure
within the $n$-particle sector, $n,\ m$ extend over the low
momentum states only and $T_{nm}$ are the partonic matrix elements
of the effective current operator. Here the helicity states of the
pion are implied in both sides. The dependence on the scale
separating low (non-perturbative) and high momenta (perturbative)
is indicated by $\mu_f$. For the valence quark state of the pion,
its light cone (LC) wave functions are defined in terms of the
bilocal operator matrix element\cite{BenekeFeldmann},
\begin{eqnarray}
 \langle \pi(p)|\bar q_\beta(z) q_\alpha|0\rangle
 &=& \frac{i f_{\pi}}{4} \int_0^1 dx\int d^2 {\bf k}_\bot
    e^{i(x p\cdot z- {\bf k}_\bot \cdot {\bf z}_\bot)} \nonumber \\
&\times& \left\{ \slash\!\!\!p \gamma_5 \psi_{\pi}(x, {\bf
k}_\bot)- \mu_\pi\gamma_5 \left( \psi_p(x, {\bf k}_\bot)
    - \sigma_{\mu\nu} p^\mu z^\nu \frac{\psi_\sigma(x,{\bf k}_\bot)}{6}
    \right) \right\}_{\alpha\beta},\label{waveoperator}
\end{eqnarray}
where $\mu_\pi=m_\pi^2/(m_u+m_d)$ and $f_{\pi}$ is the pion decay
constant, whose experimental value is $130.7\pm 0.1\pm0.36
MeV$\cite{pdg}. $\psi_{\pi}(x,\mathbf{k_{\perp}})$ is the leading
twist (twist-2) wave function, $\psi_p(x,\mathbf{k_{\perp}})$ and
$\psi_{\sigma}(x,\mathbf{k_{\perp}})$ are sub-leading twist
(twist-3) wave functions that correspond to the pseudo-scalar
structure and the pseudo-tensor structure
respectively\cite{Braun2}. The distribution amplitude $\phi(x)$
and the wave function $\psi(x,\mathbf{k_{\perp}})$ are related by
\begin{equation}\label{phipsi}
\phi(x)=\int_{|\mathbf{k}_\perp|<\mu_f}
\frac{d^2\mathbf{k_\perp}}{16\pi^3}\psi(x,\mathbf{k_\perp}).
\end{equation}

It has been shown in different approaches\cite{huangt,lis} that
applying pQCD to the pion form factor begins to be self-consistent
for a momentum transfer at about $Q^2\sim 4GeV^2$. The
next-to-leading order (NLO) QCD corrections to the pion form
factor at large momentum transfer has also been
analyzed\cite{field,grunberg,webber,stefanis,akhoury,krasnikov,bakulev,melic}.
Ref.\cite{melic} presents a complete NLO pQCD prediction for the
pion form factor and it shows that a reliable pQCD prediction can
be made at a momentum transfer around $(5-10) GeV$ with
corrections to the LO results being up to $\sim 30\%$. The
theoretical uncertainty related to the renormalization scale
ambiguity has been estimated to be less than $10\%$ and for all
the considered DAs, concerning the choices of the renormalization
schemes and the factorization scales, the ratio of the NLO to the
LO contribution to the pion form factor $F_{\pi}(Q^2)$ is greater
than $30\%$ as $Q^2<20GeV^2$.

A detailed calculation about the higher helicity components'
contributions to the hard part and the soft part of the pion form
factor within the LC pQCD approach was presented in
Ref.\cite{huangww}. Their results show that by fully keeping the
transverse momentum dependence in the hard part, the asymptotic
behavior of the hard scattering amplitude from the higher helicity
components is of order $1/Q^4$, but it can give a sizable
contribution to the pion form factor at the present experimentally
accessible energy region.

Other power corrections are from the higher twist structures in
the pion DA. In the literature, based on the asymptotic behavior
of the twist-3 DAs, especially $\phi^{as}_p(x)=1$, most of
calculations give large twist-3
contributions\cite{weiy,cdh,twist1,twist2,twist3}, i.e. the
twist-3 contribution to the pion form factor is comparable or even
larger than that of the leading twist in a wide intermediate
energy region, e.g. $Q^2\sim (2-40) GeV^2$. It is hard to believe
these results are reliable, since the power suppressed corrections
make such a large contribution up to $40GeV^2$. However, because
the end-point singularity becomes more serious, the calculations
for these higher twist contributions have more uncertainty than
that for the leading twist. In fact, one may find that such kind
of large contribution comes mainly from the end-point region and
is model dependent. It means that one should try to look for a
reasonable twist-3 wave function with a better behavior in the
end-point region than that of the asymptotic one, and the twist-3
contribution might be less important and less uncertainty.

Recently in Ref.\cite{huang3}, based on the moment calculation,
the authors obtained a new form for $\phi_p(x)$, which has a
better behavior at the end-point region than that of the
asymptotic one. Their approach is different from that of
Refs.\cite{Braun2,Ball,ball2}, i.e. they did not apply the
equation of motion for the quarks in the hadron and determined the
coefficients of the Gegenbauer polynomial expansions directly from
the DA moments obtained in the QCD sum rules. The $\phi_p(x)$
obtained in Ref.\cite{huang3} can be used to suppress the
end-point singularity coming from the hard scattering kernel. In
this paper, we will develop it to construct a model wave function
$\psi_p(x,\mathbf{k_{\perp}})$ and apply it to calculate the
twist-3 contributions to the pion form factor.

The remainder of the paper is organized as follows. In Sec.II, we
construct a model for the pionic twist-3 wave function
$\psi_p(x,\mathbf{k_{\perp}})$ with the help of the moment
calculation in Ref.\cite{huang3}. And in Sec.III, the twist-3
contribution to the pion form factor, including those coming from
the higher helicity components, will be studied within the
modified pQCD approach. In Sec.IV, we discuss the model dependence
for the twist-3 contribution. Finally we summarize our results and
give the combined hard contributions to the pion form factor in
Sec.V.

\section{A model for the pionic twist-3 wave function}

For the twist-3 DAs, since the asymptotic behavior of $\phi_p(x)$
and $\phi_\sigma(x)$ are, $\phi^{as}_p(x)\sim 1$ and
$\phi^{as}_\sigma(x)\sim 6x(1-x)$ respectively, one may observe
that the end-point singularity comes more seriously from
$\phi_p(x)$ than from $\phi_\sigma(x)$. With $\phi_\sigma(x)$ in
the asymptotic form, the end-point singularity coming from the
hard scattering kernel can be cured, while the asymptotic behavior
of $\phi_p(x)$ can not suppress such kind of end-point
singularity.

The pion twist-3 DAs have been studied in
Refs.\cite{Braun2,Ball,ball2}. They employed the conformal
symmetry and the equations of motion of the on-shell quarks within
the hadron to get the relations among the two-particle twist-3
DAs, i.e. $\phi_p(\xi)$ and $\phi_\sigma(\xi)$ (here and hereafter
$\xi\equiv (2x-1)$), and the three-particle twist-3 DA
$\phi_{3\pi}(\alpha_i)$ ($\alpha_i$ ($i=1,2,3$) is the
longitudinal momentum fraction of the corresponding constituent in
the three-particle state (higher Fock state, e.g.
$|u\bar{d}g\rangle$) of the pion and satisfies
$\sum_i\alpha_i=1$). Then they took the moments of
$\phi_{3\pi}(\alpha_i)$ to obtain the approximate forms for the
two-particle twist-3 DAs. However as has been argued in
Ref.\cite{huang3}, since the quarks are not on-shell, it is
questionable to use the equation of motion. So Ref.\cite{huang3}
suggested to calculate the moments of the pion two-particle
twist-3 DAs directly from the QCD sum rules.

Under the approximation that the lowest pole dominate and the
higher dimension condensates are negligible, the sum rule for the
moments of $\phi_p(\xi)$ can be written as\cite{huang3},
\begin{eqnarray}\label{moms}
\langle \xi_p^{2n} \rangle \cdot \langle \xi^0_p \rangle
&=&\frac{M^4}{(m_0^p)^2 f^2_{\pi}} e^{m_{\pi}^2/M^2}\left[
\frac{3}{8\pi^2}\frac{1}{2n+1} \left(
    1-(1+\frac{s_{\pi}}{M^2})e^{-\frac{s_\pi}{M^2}}\right) \right.
    -\frac{2n-1}{2}\frac{(m_u+m_d)\langle \bar{\psi}\psi \rangle}{M^4} \nonumber \\
& & +\frac{2n+3}{24} \frac{\langle
    \frac{ \alpha_s}{\pi}G^2 \rangle}{M^4} \left.  -\frac{16
    \pi}{81}\left( 21 + 8n(n+1) \right) \frac{\langle
    \sqrt{\alpha_s}\bar{\psi}\psi \rangle^2}{M^6} \right],
\end{eqnarray}
where $M$ is the Borel parameter and $\langle \xi_p^{2n}\rangle$
is the moment of $\phi_p(\xi)$, which is defined by $ \langle
\xi_p^{2n} \rangle=\frac{1}{2}\int_{-1}^{1}\xi^{2n}\phi_p(\xi)d\xi
$. The parameter $s_\pi$ in Eq.(\ref{moms}) should be chosen to
make the moments and the parameter $m_0^p$ most stable against
$M^2$ in a certain range. In Eq.(\ref{moms}), one may observe that
the usual $\mu_\pi$-dependence in the sum rule for the moments of
$\phi_p(\xi)$\cite{Braun2,Ball,ball2} has been replaced by an
undetermined parameter $m_0^p$. With the help of Eq.(\ref{moms}),
setting $\langle \xi^0_p \rangle=1$ and varying the Borel
parameter $M$ in a reasonable range, we can obtain the values for
the moments that are necessary to fit the parameters for our model
wave function.

\begin{figure}
\centering
\includegraphics[width=0.55\textwidth]{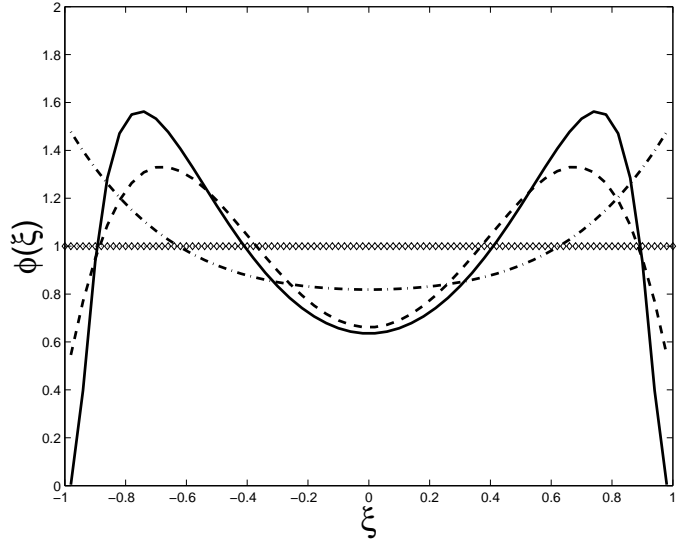}
\caption{Different type of twist-3 DA. The solid line is for our
$\phi_p(\xi)$. And for comparison, we list the asymptotic DA, the
DAs of Ref.\cite{huang3} and Refs.\cite{Braun2,ball2} in diamond
line, the dashed line and the dash-dot line respectively. }
\label{wavep1}
\end{figure}

Now we construct a model wave function
$\psi_{p}(x,\mathbf{k_\perp})$ of the twist-3 part that is related
to $\phi_p(x)$ by the definition Eq.(\ref{phipsi}). The intrinsic
transverse momentum dependence is determined by the
non-perturbative dynamics and at present we cannot solve it.
Ref.\cite{bhl} suggested a connection between the equal-time wave
function $\psi_{c.m.}(\mathbf{q}_\perp)$ in the rest frame and the
LC wave function $\psi_{LC}(x,\mathbf{k}_\perp)$ in the infinite
momentum frame, i.e.
\begin{equation}\label{bhleq}
\psi_{c.m.}(\mathbf{q}_\perp)\leftrightarrow
\psi_{LC}\left(\frac{\mathbf{k}_\perp^2+m^2}{4x(1-x)}-m^2\right),
\end{equation}
which expressed that the LC wave function should be a function of
the bound state off-shell energy. Eq.(\ref{bhleq}) is the so
called BHL prescription\cite{bhldefine}. Recently, some
improvements on the transverse momentum dependence of the wave
function have been given in Ref.\cite{qiao}, which presents a
systematic study of the B meson LC wave function in the
heavy-quark limit and by applying the QCD equations of motion.
Their results show that under the Wandzura-Wilczek approximation
\cite{Braun2,ww}, the transverse and the longitudinal momenta in
the B meson wave function are correlated through the combination
$\sim \mathbf{k}_\perp^2/x(1-x)$\footnote{Here
$x=\omega/(2\bar{\Lambda})\in(0,1)$, where $\omega$, roughly
speaking, is the longitudinal momentum of the light quark in B
meson and $\bar{\Lambda}=(M-m_b)$ is the ``effective mass" of B
meson in the heavy quark effective theory.}. By adopting the above
prescription Eq.(\ref{bhleq}) and by using the harmonic oscillator
model in the rest frame, the transverse momentum dependence part,
$\Sigma(x,\mathbf{k_\perp})$, can be written as\cite{hms},
\begin{equation} \label{sigma}
\Sigma(x,\mathbf{k_\perp})\propto \exp \left( -
 \frac{m^2+k_\perp^2}{8\beta^2x(1-x)} \right),
\end{equation}
where $m$ and $\beta$ are the quark mass and the harmonic
parameter, respectively. Combining it with the new form of
$\phi_p(\xi)$, which is in the Gegenbauer polynomial
expansion\cite{Braun2,Ball,ball2,huang3}, one can construct a
model wave function with $k_T$ dependence,
\begin{equation}\label{hel0wave}
\psi_p(x,\mathbf{k_\perp}) =(1+B_p C^{1/2}_2(1-2x)+C_p
C^{1/2}_4(1-2x))\frac{A_p}{x(1-x)}\exp \left(
-\frac{m^2+k_\perp^2}{8\beta^2x(1-x)}\right),
\end{equation}
where $C^{1/2}_2(1-2x)$ and $C^{1/2}_4(1-2x)$ are Gegenbauer
polynomials and the coefficients $A_p$, $B_p$ and $C_p$ can be
determined by the DA moments. In Eq.(\ref{hel0wave}), only the
first three terms in the Gegenbauer polynomial expansions have
been considered. Since the higher moments of $\phi_p(\xi)$
obtained from the sum rule (Eq.(\ref{moms})) depends heavily on
the Borel parameters, it is unreliable to do further expansions,
so we only take the first three moments which have a better
confidence level for our discussion. The parameters $m$ and
$\beta$ can be taken from assuming the same $k_T$ dependence as
the twist-2 wave function, and here we take\cite{hms}
\begin{equation}\label{parameter}
m=290MeV,\;\;\beta=385MeV,
\end{equation}
which are derived for $\langle\mathbf{k_\perp}^2\rangle\approx
(356MeV)^2$. From the model wave function Eq.(\ref{hel0wave}), we
obtain
\begin{equation}\label{ourphi}
\phi_p(\xi) = \frac{A_p\beta^2}{2\pi^2}(1+B_p C^{1/2}_2(\xi)+C_p
C^{1/2}_4(\xi))\exp \left( -\frac{m^2}{2\beta^2(1-\xi^2)}\right).
\end{equation}
Reasonable ranges for the $\phi_p(\xi)$ moments have been given in
Ref.\cite{huang3} by applying the QCD sum rules (Eq.(\ref{moms})),
i.e. $\langle\xi^2\rangle\sim (0.340,0.360)$ and
$\langle\xi^4\rangle\sim (0.160,0.210)$. Here we take
\begin{equation}\label{momens}
\langle\xi^0\rangle=1,\;\;\; \langle\xi^2\rangle=0.350,\;\;\;
\langle\xi^4\rangle=0.185,
\end{equation}
for our latter discussion. The parameters in the wave function can
then be determined as,
\begin{equation}
A_p=2.841\times 10^{-4}MeV^{-2},\;\;\; B_p=1.302,\;\;\; C_p=0.126.
\end{equation}
As is shown in Fig.(\ref{wavep1}), the shape of the present DA for
$\phi_p(\xi)$ is very close to the one that is proposed in
Ref.\cite{huang3}.

In the model wave function defined in Eq.(\ref{hel0wave}), only
the usual helicity components $(\lambda_1+\lambda_2=0)$ have been
taken into account, while the higher helicity components
$(\lambda_1+\lambda_2=\pm 1)$ which come from the spin-space
Wigner rotation have not been considered. As has been pointed out
in Refs.\cite{wk,huangww}, there is a large suppression coming
from the higher helicity components in the leading twist wave
function, and one may expect that the higher helicity components
in the higher twist wave functions also will do some contributions
to the pion form factor. So we need to consider the higher
helicity components in the twist-3 wave function. The full form
for the LC wave function, i.e. $\psi^{f}_p(x,\mathbf{k_\perp})$,
which includes all the helicity components, can be found in the
appendix. From $\psi^{f}_p(x,\mathbf{k_\perp})$, one may directly
find that its DA $\phi^{f}_p(\xi)$ is almost coincide with
$\phi_p(\xi)$ and for simplicity, we can take the approximate
relation, $\phi^{f}_p(\xi)\approx \phi_p(\xi)$.

In Fig.(\ref{wavep1}), we show our $\phi_p(\xi)$ in solid line,
and for comparison, we also present the asymptotic DA, the DAs of
Ref.\cite{huang3} and Refs.\cite{Braun2,ball2} in the diamond
line, the dashed line and the dash-dot line, respectively. One may
observe that the possible end-point singularity coming from the
hard scattering kernel will be suppressed in our DA and the
twist-3 contribution can be greatly suppressed at the present
experimentally accessible energy region.

\section{the twist-3 contribution to the pion
form factor in the modified pQCD approach}

In the large $Q^2$ region, by considering only the lowest valence
quark state of the pion (i.e. $n=m=2$ in Eq.(\ref{pikt})) and by
doing the Fourier transformation of the wave function with the
formula,
\begin{displaymath}
\psi(x_i,{\bf k}_{\perp};\mu_f)=\int \frac{d^{2}{\bf b}}{(2\pi)^2}
e^{-i{\bf b}\cdot {\bf k_\perp}} \hat{\psi}(x_i,{\bf b};\mu_f),
\end{displaymath}
we can transform the pion form factor Eq.(\ref{pikt}) into the
compact parameter $\mathbf{ b}$ space\cite{lis,nli},
\begin{eqnarray}
\label{twistpion} F_{\pi}(Q^2)&=&\int[dx_id{\bf b}][dy_jd{\bf h}]
\hat{\psi}(x_i,{\bf b};\mu_f)\hat{T}(x_i,{\bf b};y_j,{\bf
h};\mu_f)\hat{\psi}(y_j,{\bf h};\mu_f)
\times S_t(x_i)S_t(y_j)\times\nonumber\\
& & \exp(-S(x_i,y_j,Q,\mathbf{b},\mathbf{h};\mu_f)),
\end{eqnarray}
where $\hat{\mu}_f=\ln(\mu_f/\Lambda_{QCD})$, $[dx_id{\bf
b}]=dx_1dx_2d^2\mathbf{b} \delta(1-x_1-x_2)/(16\pi^3)$ and the
hard kernel
\begin{displaymath}
\hat{T}(x_i,{\bf b};y_j,{\bf h};\mu_f)=\int \frac{d^{2}{\bf
k}_\perp}{(2\pi)^2}\frac{d^{2}{\bf l}_\perp}{(2\pi)^2}e^{-i{\bf
b}\cdot {\bf k_\perp}-i{\bf h}\cdot {\bf l_\perp}} T(x_i,{\bf
k}_{\perp i};y_j,{\bf l}_{\perp j};\mu_f).\nonumber
\end{displaymath}
The factor $\exp(-S(x_i,y_j,Q,\mathbf{b},\mathbf{h};\mu_f))$
contains the Sudakov logarithmic corrections and the
renormalization group evolution effects of both the wave functions
and the hard scattering amplitude,
\begin{equation}
S(x_1,y_1,Q,\mathbf{b},\mathbf{h};\mu_f)= \left[\left(\sum_{i=1}^2
s(x_i,b,Q)+\sum_{j=1}^{2}s(y_j,h,Q) \right)
-\frac{1}{\beta_{1}}\ln\frac{\hat{\mu}_f}{\hat{b}}
-\frac{1}{\beta_{1}}\ln\frac{\hat{\mu}_f}{\hat{h}} \right],
\end{equation}
where ${\hat b} \equiv  {\rm ln}(1/b\Lambda_{QCD})$, ${\hat h}
\equiv {\rm ln}(1/h\Lambda_{QCD}) $ and $s(x,b,Q)$ is the Sudakov
exponent factor, whose explicit form up to next-to-leading log
approximation can be found in Ref.\cite{liyu}. In
Eq.(\ref{twistpion}), $S_t(x_i)$ and $S_t(y_i)$ come from the
threshold resummation effects and the exact form of each involves
one parameter integration\cite{kls}. In order to simplify the
numerical calculations, we take a simple parametrization proposed
in Ref.\cite{kls},
\begin{equation}
S_t(x)=\frac{2^{1+2c}\Gamma(3/2+c)}{\sqrt{\pi}\Gamma(1+c)}
[x(1-x)]^c\;,
\end{equation}
where the parameter $c$ is determined around $0.3$ for the pion
case.

To obtain the momentum projector for the pion, one may take the
Fourier transformation of the bilocal operator matrix element
defined in Eq.(\ref{waveoperator})\cite{BenekeFeldmann},
\begin{equation}
M_{\alpha\beta}^{\pi} = \frac{i f_{\pi}}{4} \Bigg\{
\slash\!\!\!p\,\gamma_5\,\psi_{\pi}(x, \mathbf{k_\perp})-
m^p_0\gamma_5 \left(\psi_p(x, \mathbf{k_\perp})
-i\sigma_{\mu\nu}\left(n^{\mu}\bar{n}^{\nu}\,\frac{\psi_{\sigma}'(x,
\mathbf{k_\perp})}{6}-p^\mu\,\frac{\psi_\sigma(x,
\mathbf{k_\perp})}{6}\, \frac{\partial}{\partial
\mathbf{k}_{\perp\nu}} \right)\right)
\Bigg\}_{\alpha\beta},\label{benek}
\end{equation}
where $\psi_{\sigma}'(x, \mathbf{k_\perp})=\partial \psi_{\sigma}
(x, \mathbf{k_\perp})/\partial x$. $n=(1,0,\mathbf{0}_\bot)$ and
$\bar{n}=(0,1,\mathbf{0}_\bot)$ are two unit vectors that point to
the plus and the minus directions, respectively. Note we have used
the parameter $m^p_0$ to replace the factor $\mu_{\pi}$ in
Eq.(\ref{benek}).

With the help of the above equations, the final formula for the
pion form factor in the modified pQCD approach can be written as,
\begin{eqnarray}\label{finalQ}
F_{\pi}(Q^2)&=&\frac{16}{9}\pi f^2_{\pi}Q^2\int_0^1
dxdy\int_0^{\infty}bdb hdh \alpha_{s}(\mu_f)\times\left[
\frac{\bar{y}}{2}\hat{\psi}_{\pi}(x,b;\mu_f)\hat{\psi}^*_{\pi}
(y,h;\mu_f)+\right.\nonumber\\
& &\frac{(m^{p}_0)^2}{Q^2}\left(y \hat{\psi}_{p}(x,b;\mu_f)
\hat{\psi}^*_{p} (y,h;\mu_f)+(1+\bar{y})
\hat{\psi}_{p}(x,b;\mu_f)\frac{\hat{\psi}^{\prime *}_{\sigma}
(y,h;\mu_f)}{6}+\right.\nonumber\\
& &\left.\left.
3\hat{\psi}_{p}(x,b;\mu_f)\frac{\hat{\psi}^*_{\sigma}
(y,h;\mu_f)}{6} \right)\right] \hat{T}(x,{\bf b};y,{\bf h};\mu_f)
\times S_t(x_i)S_t(y_j)\times\nonumber\\
& & \exp(-S(x_i,y_j,Q,\mathbf{b},\mathbf{h};\mu_f))\ ,
\end{eqnarray}
where $\bar{x}=(1-x)$, $\bar{y}=(1-y)$ and $\hat{\psi}^{\prime
*}_{\sigma} (y,h;\mu_f)=\partial \hat{\psi}^{*}_{\sigma}
(y,h;\mu_f)/\partial y$. The first term in the square bracket
gives the general twist-2 contribution and the remaining terms
that are proportional to an overall factor $((m^{p}_0)^2/Q^2)$
give the twist-3 contribution. The hard scattering amplitude
$\hat{T}(x,{\bf b};y,{\bf h};\mu_f)$ is given by
\begin{eqnarray}
\hat{T}(x,{\bf b};y,{\bf h};\mu_f)&=&
K_0\left(\sqrt{\bar{x}\bar{y}}Qb\right)\Big(\theta(b-h)K_0
\left(\sqrt{\bar{y}}Qb\right)I_0
\left(\sqrt{\bar{y}}Qh\right)+\nonumber\\
& & \theta(h-b)K_0 \left(\sqrt{\bar{y}}Qh\right)I_0
\left(\sqrt{\bar{y}}Qb\right) \Big),
\end{eqnarray}
where the higher power suppressed terms such as
$(\mathbf{k_\perp}^2/Q^2)$ has been neglected in the numerator,
$I_0$ and $K_0$ are the modified Bessel functions of the first
kind and the second kind respectively. If taking out the threshold
factors and absorbing the Sudakov factor into the definition of
the wave functions, Eq.(\ref{finalQ}) agrees with Eq.(8) in
Ref.\cite{weiy} (the factor before
$\left(\hat{\psi}_{p}(x,b;\mu_f)\hat{\psi}^*_{\sigma}
(y,h;\mu_f)/6\right)$ should be 3 other than 2 obtained there.).
To ensure that the pQCD approach is really applicable, one has to
specify carefully the renormalization scale $\mu_f$ in the strong
coupling constant. There are many equivalent ways to do so, a
popular way is to freeze $\alpha_s(Q^2)$ at lower
$Q^2$\cite{huangt,mack,parisi,curci,bjp}. Here we take the scheme
that is proposed in Refs.\cite{lis,weiy}, i.e. its value is taken
as the largest renormalization scale associated with the exchanged
virtual gluon in the longitudinal and transverse degrees,
\begin{equation}\label{scale}
\mu_f=\max(\sqrt{\bar{x}\bar{y}}Q,1/b,1/h),
\end{equation}
The Landau pole in the coupling constant at $\mu_f=\Lambda_{QCD}$
can be safely avoided in this way.

\begin{table}
\caption{The full form of the LC wave function
$\psi^{f}(x,\mathbf{k_\perp})=\psi(x,\mathbf{k_\perp})\chi_\pi$
with the helicity function $\chi_\pi$ being included.
$\psi^{f}(x,\mathbf{k_\perp})$ stands for
$\psi^{f}_\pi(x,\mathbf{k_\perp})$,
$\psi^{f}_p(x,\mathbf{k_\perp})$ and
$\psi^{f}_\sigma(x,\mathbf{k_\perp})$, respectively.}
\begin{center}
\begin{tabular}{|c||c|c|c|c|}
\hline\hline ~~~$\lambda_1\lambda_2$~~~ &
~~~$\uparrow\uparrow$~~~& ~~~$\uparrow\downarrow$~~~&
~~~$\downarrow\uparrow$~~~ & ~~~$\downarrow\downarrow$~~~ \\
\hline $\psi^{f}_{\lambda_1\lambda_2}(x,\mathbf{k_\perp})$ &
$-\frac{k_x- i k_y} {\sqrt{2(m^2+k^2)}}\psi(x,\mathbf{k_\perp})$ &
$\frac{m}{\sqrt{2(m^2+k^2)}}\psi(x,\mathbf{k_\perp})$&
$-\frac{m}{\sqrt{2(m^2+k^2)}}\psi(x,\mathbf{k_\perp})$ &
$-\frac{k_x+i k_y}{\sqrt{2(m^2+k^2)}}\psi(x,\mathbf{k_\perp})$
\\\hline\hline
\end{tabular}
\label{tab1}
\end{center}
\end{table}

Only the usual helicity components $(\lambda_1+\lambda_2=0)$ in
the pion wave function have been considered in Eq.(\ref{finalQ}).
From Eq.(\ref{lcwave}) in the appendix, one may observe that the
full form of the pion LC wave function have four helicity
components (Table. \ref{tab1}): namely,
\begin{equation}
\psi^{f}=(\psi^{f}_{\uparrow\uparrow},\psi^{f}_{\uparrow\downarrow},
\psi^{f}_{\downarrow\uparrow},\psi^{f}_{\downarrow\downarrow}),\;\;\;\;
(\psi^{f}=\psi^{f}_\pi,\;\psi^{f}_p,\;\psi^{f}_\sigma)
\end{equation}
By including the higher helicity components into the pion form
factor, Eq.(\ref{finalQ}) can be improved as
\begin{eqnarray}\label{finalQh}
F_{\pi}(Q^2)&=&\frac{16}{9}\pi f^2_{\pi}Q^2\int_0^1
dxdy\int_0^{\infty}bdb hdh \alpha_{s}(\mu_f)\times\left[
\frac{\bar{y}}{2}\sum_{\lambda_1\lambda_2}{\cal
P}(\hat\psi^{f}_{\pi},\lambda_1,\lambda_2)+\right.\nonumber\\
& &\frac{(m^p_0)^2}{Q^2}\left(y \sum_{\lambda_1\lambda_2}{\cal
P}(\hat\psi^{f}_{p},\lambda_1,\lambda_2)+
\frac{(1+\bar{y})}{6}\sum_{\lambda_1\lambda_2}{\cal
P}(\hat\psi^{f'}_{\sigma},\lambda_1,\lambda_2)+\right.\nonumber\\
& &\left.\left. \frac{1}{2}\sum_{\lambda_1\lambda_2}{\cal
P}(\hat\psi^{f}_{\sigma},\lambda_1,\lambda_2) \right)\right]
\hat{T}(x,{\bf b};y,{\bf h};\mu_f)\times S_t(x_i)S_t(y_j)\times\nonumber\\
& & \exp(-S(x_i,y_j,Q,\mathbf{b},\mathbf{h};\mu_f))\ ,
\end{eqnarray}
where $\hat\psi^{f'}_{\sigma}=\partial
\hat\psi^{f}_{\sigma}/\partial x$ and
$$\sum_{\lambda_1\lambda_2}{\cal
P}(\hat\psi^{f}_{\pi},\lambda_1,\lambda_2)=(\hat\psi^{f*}_{\pi\uparrow\downarrow}
\hat\psi^{f}_{\pi\uparrow\downarrow}
+\hat\psi^{f*}_{\pi\downarrow\uparrow}
\hat\psi^{f}_{\pi\downarrow\uparrow})-(\hat\psi^{f*}_{\pi\uparrow\uparrow}
\hat\psi^{f}_{\pi\uparrow\uparrow}+
\hat\psi^{f*}_{\pi\downarrow\downarrow}
\hat\psi^{f}_{\pi\downarrow\downarrow}),$$
$$\sum_{\lambda_1\lambda_2}{\cal
P}(\hat\psi^{f}_{p},\lambda_1,\lambda_2)=(\hat\psi^{f*}_{p\uparrow\downarrow}
\hat\psi^{f}_{p\uparrow\downarrow}
+\hat\psi^{f*}_{p\downarrow\uparrow}
\hat\psi^{f}_{p\downarrow\uparrow})-(\hat\psi^{f*}_{p\uparrow\uparrow}
\hat\psi^{f}_{p\uparrow\uparrow}+
\hat\psi^{f*}_{p\downarrow\downarrow}
\hat\psi^{f}_{p\downarrow\downarrow}),$$
$$\sum_{\lambda_1\lambda_2}{\cal
P}(\hat\psi^{f'}_{\sigma},\lambda_1,\lambda_2)=(\hat\psi^{f*}_{p\uparrow\downarrow}
\hat\psi^{f'}_{\sigma\uparrow\downarrow}
+\hat\psi^{f*}_{p\downarrow\uparrow}
\hat\psi^{f'}_{\sigma\downarrow\uparrow})-(\hat\psi^{f*}_{p\uparrow\uparrow}
\hat\psi^{f'}_{\sigma\uparrow\uparrow}+
\hat\psi^{f*}_{p\downarrow\downarrow}
\hat\psi^{f'}_{\sigma\downarrow\downarrow}),$$
$$\sum_{\lambda_1\lambda_2}{\cal
P}(\hat\psi^{f}_{\sigma},\lambda_1,\lambda_2)=(\hat\psi^{f*}_{p\uparrow\downarrow}
\hat\psi^{f}_{\sigma\uparrow\downarrow}
+\hat\psi^{f*}_{p\downarrow\uparrow}
\hat\psi^{f}_{\sigma\downarrow\uparrow})-(\hat\psi^{f*}_{p\uparrow\uparrow}
\hat\psi^{f}_{\sigma\uparrow\uparrow}+
\hat\psi^{f*}_{p\downarrow\downarrow}
\hat\psi^{f}_{\sigma\downarrow\downarrow}).$$ In the above
equation, because both the photon and the gluon are vector
particles, the quark helicity is conserved at each vertex in the
limit of vanishing quark mass\cite{lb}. Hence there is no
hard-scattering amplitude with the quark's and the antiquark's
helicities being changed. For the hard scattering amplitude
$\hat{T}(x,{\bf b};y,{\bf h};\mu_f)$, we have implicitly adopted
the approximate relation for all the twist structures in
Eq.(\ref{finalQh}), i.e.
\begin{equation}\label{apphel}
\hat{T}(x,{\bf b};y,{\bf
h};\mu_f)^{\uparrow\downarrow+\downarrow\uparrow}\approx -
\hat{T}(x,{\bf b};y,{\bf
h};\mu_f)^{\uparrow\uparrow+\downarrow\downarrow}.
\end{equation}
By ignoring the transverse momentum dependence in the quark
propagator and applying the symmetries of the wave functions,
especially the fact that
$\psi^{f*}_{\uparrow\uparrow}(x,\mathbf{k_\perp})=
\psi^{f}_{\downarrow\downarrow}(x,\mathbf{k_\perp})$,
Ref.\cite{wk} pointed out that the approximate relation
Eq.(\ref{apphel}) can be strictly satisfied. In fact, when the
transverse momentum dependence in the quark propagator has been
ignored, the $T_H$ depends only on one compact $\mathbf{b}$-space,
and Eq.(\ref{apphel}) can be changed to a strict one, i.e.
$\hat{T}(x,y,{\bf
b};\mu_f)^{\uparrow\downarrow+\downarrow\uparrow}= -
\hat{T}(x,y,{\bf b};\mu_f)^{\uparrow\uparrow+
\downarrow\downarrow}$. As is shown in Ref.\cite{lis}, the
transverse momentum dependence in the quark propagator will give
about $15\%$ correction at $Q=2GeV$\cite{lis}, so this effect can
not be safely neglected. The hard scattering amplitude for the
twist-2 contribution has been strictly calculated in
Ref.\cite{huangww} within the LC pQCD approach. One may find that
when all the $k_T$ dependence are included, strictly
$\hat{T}(x,{\bf b};y,{\bf
h};\mu_f)^{\uparrow\downarrow+\downarrow\uparrow}\neq -
\hat{T}(x,{\bf b};y,{\bf
h};\mu_f)^{\uparrow\uparrow+\downarrow\downarrow}$ and
Eq.(\ref{apphel}) can be approximately satisfied. In the following
discussions, we will keep the transverse momentum dependence in
the hard scattering amplitude fully and use the approximate
relation Eq.(\ref{apphel}) to estimate all the helicity
components' contributions to the pion form factor.

Before doing numerical calculations, we would like to mention a
few words on the value of $m^p_0$. Based on the equation of motion
of on-shell quarks, the authors used
$\mu_{\pi}=m_{\pi}^2/(m_u+m_d)\sim 2.0GeV$ instead of $m^p_0$ for
the twist-3 wave functions in Refs.\cite{Braun2,beneke,weiy}. A
running behavior has been introduced in
Refs.\cite{twist1,twist2,twist3,cdh} and with this choice, one may
find that the average value for $\mu_{\pi}$ over the intermediate
energy region is around $2.5GeV$. In Refs.\cite{kls,ly} a smaller
phenomenological value $\sim 1.4GeV$, which is consistent with the
result obtained from the chiral perturbation
theory\cite{chiral,ball2}, is used to fit the B meson to the light
meson transition form factors. Based on the moment calculation by
applying the QCD sum rules, Ref.\cite{huang3} obtained
$m^p_0=1.30\pm0.06GeV$, which is very close to the above
phenomenological value. So to be consistent with our model wave
function constructed in the last section, we will take
$m^p_0=1.30GeV$ for our latter discussions.

\begin{figure}
\centering
\includegraphics[width=0.55\textwidth]{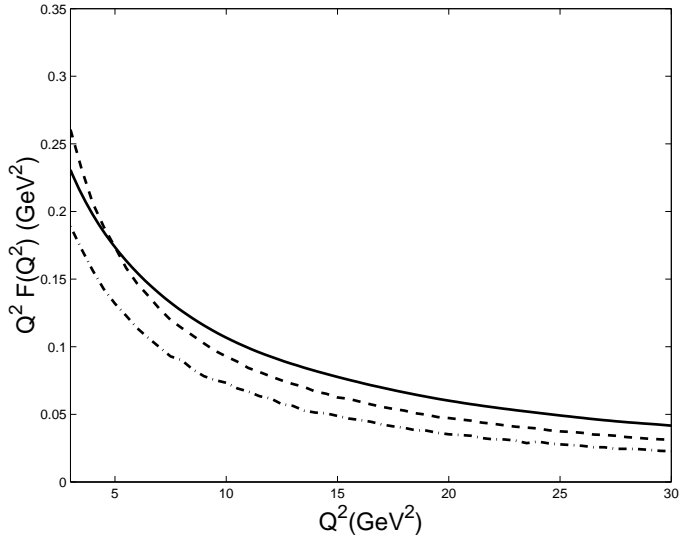}
\caption{Twist-3 contribution to the pion form factor $Q^2F(Q^2)$,
where the second moment of $\psi_p(x,\mathbf{k}_\perp)$ or
$\psi^f_p(x,\mathbf{k}_\perp)$ is taken to be
$\langle\xi^2\rangle=0.350$. The dash-dot line and the dashed line
are the twist-3 contributions for the full form of the LC wave
function with or without considering the $k_T$ dependence in the
quark propagator. The solid line is for the twist-3 contribution
from the LC wave function that contains only the usual helicity
component but is normalized to unity.} \label{fig:3}
\end{figure}

We show the twist-3 contribution to pion form factor $Q^2F(Q^2)$
with all helicity components (i.e. using the full form of the LC
wave functions $\psi^f_p(x,\mathbf{k_\perp})$ and
$\psi^f_\sigma(x,\mathbf{k_\perp})$) calculated within the
modified pQCD approach in Fig.(\ref{fig:3}), where the second
moment of $\psi^f_p(x,\mathbf{k}_\perp)$ is taken to be
$\langle\xi^2\rangle=0.350$. One may observe that the transverse
momentum dependence in the quark propagator will give about $25\%$
correction at $Q^2=2GeV$ for the twist-3 contribution, which is
bigger than the case of the leading twist contribution. So it is
more essential to keep the transverse momentum dependence fully
into the hard scattering kernel for the twist-3 contribution. As a
comparison, we also show the contribution from the twist-3 wave
functions (i.e. $\psi_p(x,\mathbf{k_\perp})$ and
$\psi_\sigma(x,\mathbf{k_\perp})$) that contain only the usual
helicity component but are normalized to unity in
Fig.(\ref{fig:3}). One may find the contribution from the twist-3
wave function that contains only the usual helicity component but
is normalized to unity (the solid line) is larger than the
contribution from the wave function with all the helicity
components being considered (the dash-dot line). It is reasonable
and is also the case of the twist-2 contribuion\cite{huangww},
because if one normalizes the valence Fock state to unity without
including the higher helicity components, then the contribution
from the valence state can be enhanced and become important
inadequately.

\section{Comparison with other models for twist-3 wave function}

As has been pointed out in Sec.III, the contribution from the
twist-3 wave functions $\psi_p(x,\mathbf{k_\perp})$ and
$\psi_\sigma(x,\mathbf{k_\perp})$ that contain only the usual
helicity components but is normalized to unity is larger than the
contribution from the wave functions
$\psi^f_p(x,\mathbf{k_\perp})$ and
$\psi^f_\sigma(x,\mathbf{k_\perp})$ with all the helicity
components being considered. However, as is shown in
Fig.(\ref{fig:3}), since both of the twist-3 contributions have a
similar behavior and are close to each other, the qualitative
conclusions will be the same. And for easy comparing with the
results in the literature, we will take the LC wave functions
$\psi_p(x,\mathbf{k_\perp})$ and $\psi_\sigma(x,\mathbf{k_\perp})$
that only contain the usual helicity components for the
discussions in the present section.

Because of the end-point singularity, the twist-3 contribution
depends heavily on the twist-3 wave function, especially on
$\psi_p(x,\mathbf{k_\perp})$. In this section, we will do a
comparative study on the twist-3 contribution from different type
of $\psi_p(x,\mathbf{k_\perp})$. For this purpose, we take
Eq.(\ref{finalQ}) to calculate the pion form factor, in which only
the usual helicity components in the wave functions have been
taken into consideration.

The twist-2 and twist-3 wave functions $\psi_{\pi}$, $\psi_p$ and
$\psi_{\sigma}$ may have different transverse momentum dependence,
and for simplicity, we assume the same transverse momentum
dependence for these space wave functions. For the transverse
momentum dependence of the wave function, we take a simple
Gaussian form, i.e.
\begin{equation} \label{eq:sigma}
\Sigma(x,\mathbf{k_\perp})=\frac{A}{g(x)}\exp \left( -
 \frac{m^2+k_\perp^2}{8\beta^2g(x)} \right),
\end{equation}
where $A$ is the normalization factor, $g(x)$ is either $1$ or
$x(1-x)$. When $g(x)=x(1-x)$, it is agree with the BHL
prescription mentioned in Sec.II. After making the Fourier
transformation, Eq.(\ref{eq:sigma}) can be transformed into the
compact parameter $\mathbf{b}$ space as,
\begin{equation}
\Sigma(x,\mathbf{b}) = \frac{2\pi A}{g(x)}\int_0^{1/b} \exp \left(
- \frac{m^2+k^2_\perp}{8\beta^2g(x)} \right)
J_{0}(bk_\perp)k_\perp dk_\perp \ ,
\end{equation}
where the upper limit $(1/b)$ is necessary to insure the wave
function to be ``soft"\cite{botts,ch1}.

Next, we consider the pion wave functions. The twist-2 wave
function $\psi_{\pi}(x,\mathbf{k_\perp})$ with the prescription
Eq.(\ref{bhleq}) can be written as
\begin{equation}
\psi_{\pi}(x,\mathbf{k_\perp})=A_{\pi} \exp
\left(-\frac{m^2+k_\perp^2}{8\beta^2 x(1-x)} \right),
\end{equation}
where the parameters can be determined by the normalization
condition of the wave function\cite{BenekeFeldmann}
\begin{equation}\label{normalization}
\int^1_0 dx \int \frac{d^{2}{\bf
k}_{\perp}}{16\pi^3}\psi_\pi(x,{\bf k}_{\perp}) =1\ ,
\end{equation}
and some necessary constraints\cite{hms}. Taking the parameter
values in Eq.(\ref{parameter}), we obtain $A_{\pi}=1.187\times
10^{-3}MeV^{-2}$. The asymptotic form of twist-3 DA
$\phi_{\sigma}(x)$ is the same as that of $\phi_{\pi}(x)$, and the
end-point singularity coming from the hard scattering amplitude
can also be cured. So for $\phi_{\sigma}(x)$ we also take its
asymptotic form. For the twist-3 contribution to the pion form
factor, the main difference for the existed
results\cite{weiy,cdh,twist1,twist2,twist3} comes mainly from the
different models for $\psi_p(x,\mathbf{k_\perp})$. The difference
caused by the different model of
$\psi_{\sigma}(x,\mathbf{k_\perp})$ (if all of them are asymptotic
like) are quite small, so in the following, we will only consider
the difference caused by different type of
$\psi_p(x,\mathbf{k_\perp})$ and the contribution from
$\psi_{\sigma}(x,\mathbf{k_\perp})$ will be included as a default
with the fixed asymptotic form for its DA and the same $k_T$
dependence as $\psi_p(x,\mathbf{k_\perp})$.

In the asymptotic limit, $\phi^{as}_p(x)=1$, the end-point
singularity coming from the hard scattering amplitude can not be
cured, and the model dependence of $\phi_p(x)$ is much more
involved. Taking the asymptotic DA and ignoring the $k_T$
dependence in the wave function, Refs.\cite{cdh,twist1,twist2}
obtained a much larger contribution in a wide energy region
$2GeV^2<Q^2<40GeV^2$, comparing with the twist-2 contribution.
Using the model wave function of $\psi_p(x,\mathbf{k_\perp})$
constructed in Sec.II, one may find that the twist-3 contributions
are suppressed certainly.

\begin{table}
\caption{The first three moments for the twist-2 and the twist-3
wave functions, where all the full form of LC wave functions have
the same BHL-like $k_T$ dependence,
$\psi^f_{\pi}$($\psi^f_{\sigma})=
\psi_{\pi}$($\psi_{\sigma})\chi_\pi$ and
$\psi^{f}_{p}=\psi_p\chi_\pi$.}
\begin{center}
\begin{tabular}{|c||c||c||c|c|c||c||c||}
\hline\hline ~~~-~~~ & \multicolumn{5}{|c||}{~~~without Wigner
rotation~~~} & \multicolumn{2}{|c||}{~~~with Wigner rotation~~~} \\
\hline ~~~-~~~ & ~~~$\psi_{\pi}$/$\psi_{\sigma}$~~~
&~~~$\psi_{p}$~~~ & ~~~$\psi^{(1)}_{p}$~~~& ~~~$\psi^{(2)}_{p}$~~~
& ~~~$\psi^{(3)}_{p}$~~~
& ~~~$ \psi^{f}_{\pi}$/$\psi^{f}_{\sigma}$~~~ & ~~~$\psi^{f}_{p}$~~~\\
\hline\hline $\langle\xi^0\rangle$ & 1& 1&  1& 1& 1& 1& 1\\
\hline $\langle\xi^2\rangle$ &0.167 & 0.350 & 0.333& 0.391& 0.352 & 0.176 & 0.350\\
\hline $\langle\xi^4\rangle$ &0.060 & 0.185 & 0.200& 0.251& 0.197 & 0.066& 0.185\\
\hline\hline
\end{tabular}
\label{tab2}
\end{center}
\end{table}

To study this effects more clearly, we compare our model with
three different types of $\psi_p(x,\mathbf{k_\perp})$. In the
literature, most of the calculations on the twist-3 contribution
of the pion take $\phi_p(x)$ as $\psi_p(x,\mathbf{k_\perp})$, i.e.
without considering the intrinsic $k_T$ dependence in the wave
function, some examples for the electromagnetic pion form factor
can be found in Refs.\cite{cdh,twist1,twist2} and examples for the
$B\to \pi$ form factor can be found in Refs.\cite{ly,kls}.
However, as has been argued in several
papers\cite{huangww,weiy,jk}, the intrinsic transverse momentum
dependence in the wave function is very important for the pion
form factor and the results will be overestimated without
including this effect. So in our comparison, the three different
type of wave functions are constructed by adding a common simple
Gaussian form (Eq.(\ref{eq:sigma}) with $g(x)=1$) to three
different type of distribution amplitudes used in the
literature\footnote{By using the prescription Eq.(\ref{bhleq}) for
the intrinsic $k_T$ dependence (Eq.(\ref{eq:sigma}) with
$g(x)=x(1-x)$), we can construct another three different model
wave functions for $\psi_p(x,\mathbf{k_\perp})$. However one may
find that the moments of these three wave functions are too small
and are out of the reasonable range obtained from the QCD sum
rule, so we will not take them for our study.}, i.e. the one of
asymptotic behavior, the one in Ref.\cite{Braun2} and the one in
Ref.\cite{huang3} respectively,
\begin{eqnarray}\label{twopsip}
\psi^{(1)}_p(x,\mathbf{k_\perp})&=& A'_p \exp
\left( -\frac{k_\perp^2}{8\beta^{'2}} \right),\\
\psi^{(2)}_p(x,\mathbf{k_\perp})&=&
(1+0.43C_2^{1/2}(2x-1)+0.09C_4^{1/2}(2x-1))A'_p\exp \left(
-\frac{k_\perp^2}{8\beta^{'2}} \right),\\
\psi^{(3)}_p(x,\mathbf{k}_\perp)&=&
(1+0.137C_2^{1/2}(2x-1)-0.721C_4^{1/2}(2x-1)) A'_p\exp \left(
-\frac{k^2_\perp}{8\beta^{'2}} \right).
\end{eqnarray}
The parameters $A'_p$ and $\beta'$ can be determined from the
similar wave function normalization condition as
Eq.(\ref{normalization}), $A'_p=7.025\times 10^{-4}MeV^{-2}$ and
$\beta'=168MeV$. For the wave functions $\psi^{(i)}_p$
($i=1,2,3$), the harmonic parameter $\beta'$ is different from
that of $\psi_{\pi}(x,\mathbf{k_\perp})$, however it can be taken
as an effective/average value of the harmonic parameter with $m=0$
and $g(x)=1$. The moments of the corresponding DAs are listed in
Table.\ref{tab2}.

\begin{figure}
\centering
\begin{minipage}[c]{0.48\textwidth}
\centering
\includegraphics[width=2.9in]{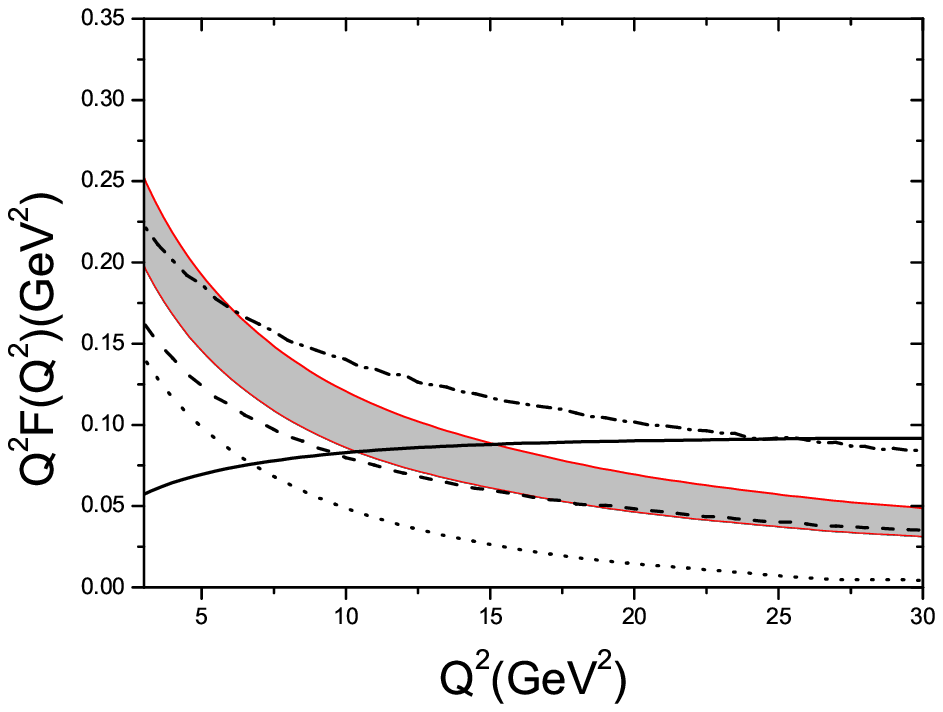}
(a)
\end{minipage}%
\hspace{0.2in}
\begin{minipage}[c]{0.48\textwidth}
\centering
\includegraphics[width=2.9in]{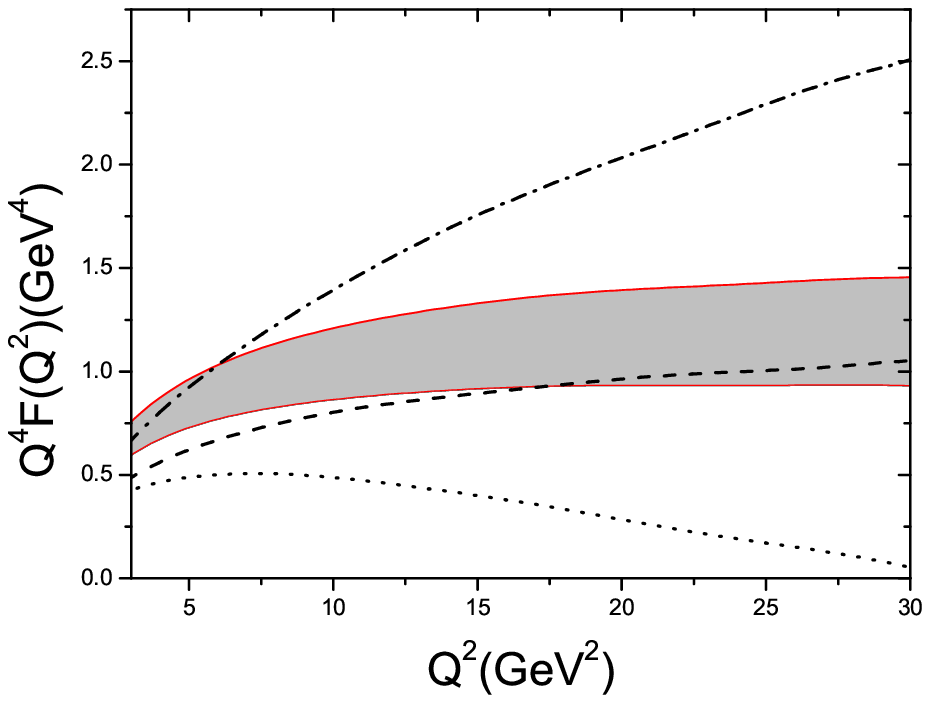}
(b)
\end{minipage}%
\caption{Different twist-3 wave function's contribution to the
pion form factor, where the left diagram is for $Q^2F(Q^2)$ and
the right is for $Q^4F(Q^2)$. The dashed line, the dash-dot line
and the dotted line are for $\psi^{(1)}_p(x,\mathbf{k_\perp})$,
$\psi^{(2)}_p(x,\mathbf{k_\perp})$ and
$\psi^{(3)}_p(x,\mathbf{k_\perp})$ respectively. The contribution
from our model wave function $\psi_p(x,\mathbf{k_\perp})$ with
varying second moment $\langle\xi^2\rangle$ is shown by a shaded
band, whose lower and upper edges correspond to
$\langle\xi^2\rangle=0.320,0.370$ respectively. For comparison,
the twist-2 contribution from $\psi_\pi(x,\mathbf{k_\perp})$ is
shown in solid line.} \label{twist3-1}
\end{figure}

We show the contributions to the pion form factor from the
different model for $\psi_p(x,\mathbf{k_\perp})$ in
Figs.(\ref{twist3-1}a,\ref{twist3-1}b), where the contribution
from our model wave function $\psi_p(x,\mathbf{k_\perp})$ with
varying second moment $\langle\xi^2\rangle$ is shown by a shaded
band and the twist-2 contribution from
$\psi_\pi(x,\mathbf{k_\perp})$ is included in
Fig.(\ref{twist3-1}a) for comparison. Our present result for
$\psi^{(1)}_p(x,\mathbf{k_\perp})$ (in dashed line) is much lower
than the result shown in Ref.\cite{weiy}, since the value of
$\mu_\pi$ used there has been changed to the present value of
$m^p_0$. One may observe that the twist-3 contribution is improved
with our model wave function, and for the case of
$\langle\xi^2\rangle=0.350$, at about $Q^2=30GeV^2$, it is only
about $45\%$ comparing with the twist-2 contribution. This
behavior is quite different from the previous
observations\cite{weiy,cdh,twist1,twist2,twist3}, where they
concluded that the twist-3 contribution to the pion form factor is
comparable or even larger than that of the leading twist in a wide
intermediate energy region.

As is shown in Figs.(\ref{twist3-1}a,\ref{twist3-1}b), the twist-3
contribution from $\psi^{(1)}_p(x,\mathbf{k_\perp})$ is comparable
to our model wave function, which also has the right power
behavior. We take a simple Gaussian behavior (Eq.(\ref{eq:sigma})
with $g(x)=1$) for the transverse momentum dependence in
$\psi^{(1)}_p(x,\mathbf{k_\perp})$, i.e. a complete factorization
between longitudinal and transverse momentum-dependence in the
wave function. This Gaussian distribution behavior shows a strong
dumping at large transverse distances, $\sim \exp(-2\beta^2 b^2)$,
while our model function with the prescription Eq.(\ref{bhleq})
has a slow-dumping with oscillatory behavior, $\sim
\cos\left(\sqrt{x(1-x)}b\beta^2-\pi/4\right) /\sqrt{b}$. If we
also take the simple transverse momentum behavior in our model
wave function, i.e. the one as $\psi_p^{(3)}(x,\mathbf{k_\perp})$,
we find that the twist-3 contribution will be even lower, which is
shown clearly by the dotted line in
Figs.(\ref{twist3-1}a,\ref{twist3-1}b). However as is shown in
Fig.(\ref{twist3-1}b), we can not achieve a right power behavior
with $\psi_p^{(3)}(x,\mathbf{k_\perp})$, i.e. it drops down too
quickly.

Finally, with our model wave function for $\psi_p(x,\mathbf{k})$,
we discuss the model dependence of the twist-3 contribution on the
DA moments $\langle\xi^{2n}\rangle$. Here we take the second
moment $\langle\xi^{2}\rangle$, which gives the main contribution
to $\phi_p(\xi)$, as an example. Varying the second moment
$\langle\xi^2\rangle$ within a broader range, i.e.
$\langle\xi^2\rangle\in (0.320,0.370)$, and adjusting the fourth
moment $\langle\xi^4\rangle$ to make $\phi_p(\xi)$ has a closed
behavior as the one that is obtained in Ref.\cite{huang3} (i.e.
the dashed line in Fig.(\ref{wavep1})), we can determine the
corresponding parameters $A_p$, $B_p$ and $C_p$ in the wave
function $\psi_p(x,\mathbf{k})$. The twist-3 contribution to the
pion form factor with varying second moment $\langle\xi^2\rangle$
has been shown by a shaded band in
Figs.(\ref{twist3-1}a,\ref{twist3-1}b). One may observe that the
pionic twist-3 contribution increases with the increment of
$\langle\xi^2\rangle$ and all has a quite similar behavior on the
variation of the energy scale $Q^2$, i.e. as is shown in
Fig.(\ref{twist3-1}b), the right asymptotic power behavior of
order $1/Q^4$ has already been achieved at the present
experimentally accessible energy region.

\section{summary and discussion}

In this paper, we have constructed a model wave function for
$\psi_p(x,\mathbf{k_\perp})$ based on the moment
calculation\cite{huang3} by using the QCD sum rule approach. It
has a better end-point behavior than that of the asymptotic one
and its moments are consistent with the QCD sum rule results.
Although its moments are slightly different from that of the
asymptotic DA, its better end-point behavior will cure the
end-point singularity of the hard scattering amplitude and its
contribution will not be overestimated at all.

With this model wave function, by keeping the $k_T$ dependence in
the wave function and taking the Sudakov effects and the threshold
effects into account, we have carefully studied the twist-3
contributions to the pion form factor. Comparing the different
models for $\psi_p(x,\mathbf{k_\perp})$, a detailed study on the
twist-3 contribution to the pion form factor has been given within
the modified pQCD approach. It has been shown that our model wave
function $\psi_p(x,\mathbf{k_\perp})$ can give the right power
behavior for the twist-3 contribution. With the present model wave
function defined in Eq.(\ref{hel0wave}) for
$\psi_p(x,\mathbf{k_\perp})$, our results predict that, at about
$Q^2\sim 10GeV^2$, the twist-3 contribution begins to be less than
the twist-2 contribution, and for the wave function
$\psi_p(x,\mathbf{k}_\perp)$ with $\langle\xi^2\rangle=0.350$ at
about $Q^2=30GeV^2$, it is only about $45\%$ comparing with the
twist-2 contribution. This behavior is quite different from the
previous observations\cite{weiy,cdh,twist1,twist2,twist3}, where
they concluded that the twist-3 contribution to the pion form
factor is comparable or even larger than that of the leading twist
in a wide intermediate energy region up to $40GeV^2$. The higher
helicity components $(\lambda_1+\lambda_2=\pm 1)$ in the twist-3
wave function that come from the spin-space Wigner rotation have
also been considered. The higher helicity components in the
twist-3 wave function will do a further suppression to the
contribution from the usual helicity components
$(\lambda_1+\lambda_2=0)$, and at about $Q^2=5GeV^2$, it will give
$\sim 10\%$ suppression.

\begin{figure}
\centering
\includegraphics[width=0.55\textwidth]{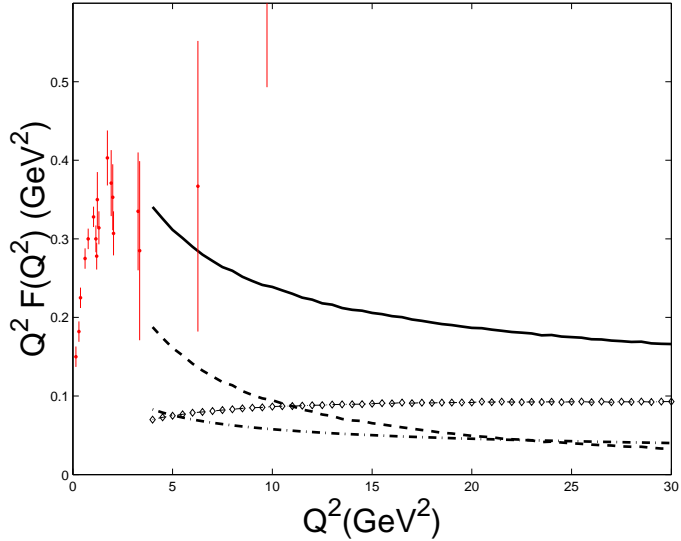}
\caption{Perturbative prediction for the pion form factor. The
diamond line, the dash-dot line, the dashed line and the solid
line are for LO twist-2 contribution, the approximate NLO twist-2
contribution\cite{field,melic}, the twist-3 contribution and the
combined total hard contribution, respectively. The experimental
data are taken from\cite{cjb}.}\label{twist3-tot}
\end{figure}

In Fig.(\ref{twist3-tot}), we show the combined hard contributions
for the twist-2 and twist-3 contributions to the pion form factor,
where the higher helicity components have been included in both
the twist-2 and the twit-3 wave functions, and the twist-3
contribution has been calculated with our model wave function
$\psi^f_p(x,\mathbf{k_\perp})$ with $\langle\xi^2\rangle=0.350$.
As has been pointed out in Refs.\cite{huangt,lis}, the
applicability of pQCD to the pion form factor can only be achieved
at a momentum transfer bigger than $Q^2\sim 4GeV^2$, so in
Fig.(\ref{twist3-tot}), all the curves are started at
$Q^2=4GeV^2$. Together with the NLO corrections to the twist-2
contributions, which for the asymptotic DA, with the
renormalization scale $\mu_R$ and the factorization scale $\mu_f$
taken to be $\mu^2_R=\mu^2_f=Q^2$, can roughly be taken
as\cite{field,melic},
$Q^2F^{NLO}_{\pi}\approx(0.903GeV^2)\alpha^2_s(Q^2)$, one may find
that the combined total hard contribution do not exceed and will
reach the present experimental data. There is still a room for the
other power corrections, such as the higher Fock states'
contributions\cite{bakker,melo}, soft contributions etc.. Finally,
we will conclude that there is no any problem with applying the
pQCD theory including all power corrections to the exclusive
processes at $Q^2>$ a few $GeV^2$.

\begin{center}
\section*{Acknowledgements}
\end{center}

This work was supported in part by the Natural Science Foundation
of China (NSFC). And the authors would like to thank M.Z. Zhou and
X.H. Wu for some useful discussions.\\

\appendix
\section{full form for the LC wave function}

By doing the spin-space Wigner rotation, we can transform the
ordinary equal-time (instant-form) spin-space wave function in the
rest frame into that in the LC dynamics. After doing the Wigner
rotation, the covariant form for the pion helicity functions can
be written as\cite{hms,jaus},
\begin{equation}
\chi_{\pi}(x,\mathbf{k_\perp})=\frac{1}{\sqrt{2}\tilde{M}_0}\bar
u(p_1,\lambda_1)\gamma_5 v(p_2,\lambda_2),
\end{equation}
where $p_1=(x,\mathbf{k_\perp})$ and
$p_2=(\bar{x},-\mathbf{k_\perp})$ ($\bar{x}=1-x$) are the momenta
of the two constituent quarks in the pion,
$\tilde{M}_0^2=(m^2+k^2)/x(1-x)$ and the LC spinors $u$ and $v$
have the Wigner rotation built into them. Then the full form of
the LC wave function can be written as
\begin{equation}\label{lcwave}
\psi^{f}(x,\mathbf{k_\perp})=\psi(x,\mathbf{k_\perp})\chi_{\pi}(x,\mathbf{k_\perp}),
\end{equation}
where the momentum space wave function $\psi(x,\mathbf{k_\perp})$
represents $\psi_p(x,\mathbf{k_\perp})$,
$\psi_\pi(x,\mathbf{k_\perp})$ and
$\psi_\sigma(x,\mathbf{k_\perp})$ respectively. Because all the LC
wave functions can be dealt with in a similar way, here we only
take $\psi_p(x,\mathbf{k_\perp})$ that is defined in
Eq.(\ref{hel0wave}) as an explicit example to show how to
determine the parameters in the full form.

The full form of LC wave function $\psi^{f}_p(x,\mathbf{k_\perp})$
contains all the helicity components' contributions and its four
components can be found in Table.\ref{tab1}. The parameter values
built in the wave function $\psi^f_p(x,\mathbf{k_\perp})$ can be
done in a similar way as for the wave function of
$\psi_p(x,\mathbf{k_\perp})$ that contains only the usual helicity
components, i.e.
\begin{eqnarray}
A_{p} = 4.088\times 10^{-4} MeV^{-2},\;\;
B_p=1.077,\;\;C_p=-4.317\times 10^{-3} \\
m=309.6MeV,\;\;\; \beta=395.9MeV,\;\;\;{\rm for}\;\;\;
\langle\mathbf{k_\perp}^2\rangle\approx(367MeV)^2
\end{eqnarray}
where the parameters $m$, $\beta$ are determined by the wave
function normalization condition and some necessary
constraints\cite{hms}, and the values of $A_p$, $B_p$ and $C_p$
are determined by requiring the first three moments of its DA to
be the values shown in Eq.(\ref{momens}). From the wave function
Eq.(\ref{lcwave}), we obtain
\begin{equation}
\phi^{f}_p(\xi) = \frac{A_p m
\beta}{\pi^{3/2}\sqrt{2(1-\xi^2)}}(1+B_p C^{1/2}_2(\xi)+C_p
C^{1/2}_4(\xi))\exp \left(1
-Erf\left(\sqrt{\frac{m^2}{2\beta^2(1-\xi^2)}}\right)\right),
\end{equation}
where the error function $Erf(x)$ is defined as $Erf(x)=
\frac{2}{\sqrt{\pi}}\int_0^x e^{-t^2}dt$. One may find that
$\phi^{f}_p(\xi)$ is almost coincide with $\phi_p(\xi)$ that is
shown in Eq.(\ref{ourphi}), and for simplicity, we can take the
approximate relation, $\phi^{f}_p(\xi)\approx \phi_p(\xi)$. It is
reasonable because we have adjusted the parameters in both DAs to
have the same moments and due the fact that the momentum space
wave function $\psi_p(x,\mathbf{k_\perp})$ is an even function of
$\mathbf{k_\perp}$, one may find that the higher helicity
components in $\psi^{f}_p(x,\mathbf{k_\perp})$ do not contribute
to $\phi^{f}_p(\xi)$.

\end{document}